\documentclass[12pt,preprint]{aastex}

\shorttitle{1.2 Meter Shielded Cassegrain Antenna}
\shortauthors{Koch et al.}

\begin{document}


\title{1.2 Meter Shielded Cassegrain Antenna for Close-Packed Radio Interferometer}


\author{
 Patrick M. Koch\altaffilmark{1},
 Philippe Raffin\altaffilmark{1},
 Yau-De Huang\altaffilmark{1},
 Ming-Tang Chen\altaffilmark{1},
 Chih-Chiang Han\altaffilmark{1},
 Kai-Yang Lin\altaffilmark{1},
 Pablo Altamirano\altaffilmark{1}, 
 Christophe Granet\altaffilmark{2}, 
 Paul T.P.Ho\altaffilmark{1,3},
 Chih-Wei L.Huang\altaffilmark{4},
 Michael Kesteven\altaffilmark{5},
 Chao-Te Li\altaffilmark{1},
 Yu-Wei Liao\altaffilmark{1,4},
 Guo-Chin Liu\altaffilmark{1,6},
 Hiroaki Nishioka\altaffilmark{1},
 Ching-Long Ong\altaffilmark{7},
 Peter Oshiro\altaffilmark{1},
 Keiichi Umetsu\altaffilmark{1},
 Fu-Cheng Wang\altaffilmark{4} \&
 Jiun-Huei Proty Wu\altaffilmark{4}
}

\altaffiltext{1}{Academia Sinica, Institute of Astronomy and
 Astrophysics, P.O.Box 23-141, Taipei 10617, Taiwan} 
\altaffiltext{2}{CSIRO ICT Centre, Epping NSW 1710, Australia}
\altaffiltext{3}{Harvard-Smithsonian Center for Astrophysics, 60
 Garden Street, Cambridge, MA 02138, USA}
\altaffiltext{4}{Department of Physics, Institute of Astrophysics, \& Center
for Theoretical Sciences, National Taiwan University, Taipei 10617, Taiwan}
\altaffiltext{5}{Australia Telescope National Facility, P.O.Box 76, Epping NSW 
 1710, Australia}
\altaffiltext{6}{Department of Physics, Tamkang University, 251-37 Tamsui,
 Taipei County, Taiwan} 
\altaffiltext{7}{CoTec Inc., Taichung, Taiwan}
\email{pmkoch@asiaa.sinica.edu.tw}
%



\begin{abstract}
Interferometric millimeter observations of the cosmic microwave 
background and clusters of galaxies with arcmin resolutions
require antenna arrays with short spacings. 
Having all antennas co-mounted on a single steerable platform sets
limits to the overall weight. A 25~kg
lightweight novel carbon-fiber
design for a 1.2~m diameter Cassegrain antenna is presented.
The finite element analysis predicts excellent structural
behavior under gravity, wind and thermal load. 
The primary and secondary mirror surfaces are aluminum coated 
with a thin TiO$_2$ top layer for protection.
A low beam 
sidelobe level is achieved with a Gaussian feed illumination 
pattern with edge taper, designed based on feedhorn antenna simulations
and verified in a far field beam pattern measurement. A shielding
baffle reduces inter-antenna coupling
to below $\sim$ -135~dB.
The overall
antenna efficiency, including a series of efficiency factors, 
is estimated to be around 60\%, with major losses coming
from the feed spillover and secondary blocking.
With this new antenna, a detection rate of about 50 clusters
per year is anticipated in a 13-element array operation.
\end{abstract}





\section{Introduction}

The Array for Microwave Background Anisotropy (AMiBA) is a forefront
radio interferometer for research in cosmology. This project is led, designed, 
constructed, and operated by the Academia Sinica,
 Institute of Astronomy and Astrophysics (ASIAA), Taiwan,  
with major collaborations with National Taiwan University, 
Physics Department (NTUP), Electrical Engineering Department (NTUEE), 
and the Australian Telescope 
National Facility (ATNF). Contributions also came from the 
Carnegie Mellon University (CMU), and the 
National Radio Astronomy Observatory (NRAO). 
As a dual-channel 86-102 GHz interferometer array of up to 19 elements, 
AMiBA is designed to have full polarization capabilities, 
sampling structures on the sky greater than 2 arcmin in size. 
The AMiBA target science is the distribution of high
red-shift clusters of galaxies via the Sunyaev-Zel'dovich Effect (SZE),
e.g. \citet{sz72,bi99,carl02}
and references therein, 
as a means to probe the primordial and early structure of the Universe. 
AMiBA will also measure the 
Cosmic Microwave Background (CMB), 
e.g. \citet{spergel07,aghanim08,larson10}, 
temperature anisotropies on scales, which are sensitive to 
structure formation scenarios of the Universe.
AMiBA is sited on Mauna Loa in Hawaii, at an elevation of 3,400m 
to take advantage of higher atmospheric transparency and minimum radio frequency interference. 

\begin{figure}
\begin{center}
\includegraphics[scale=0.65]{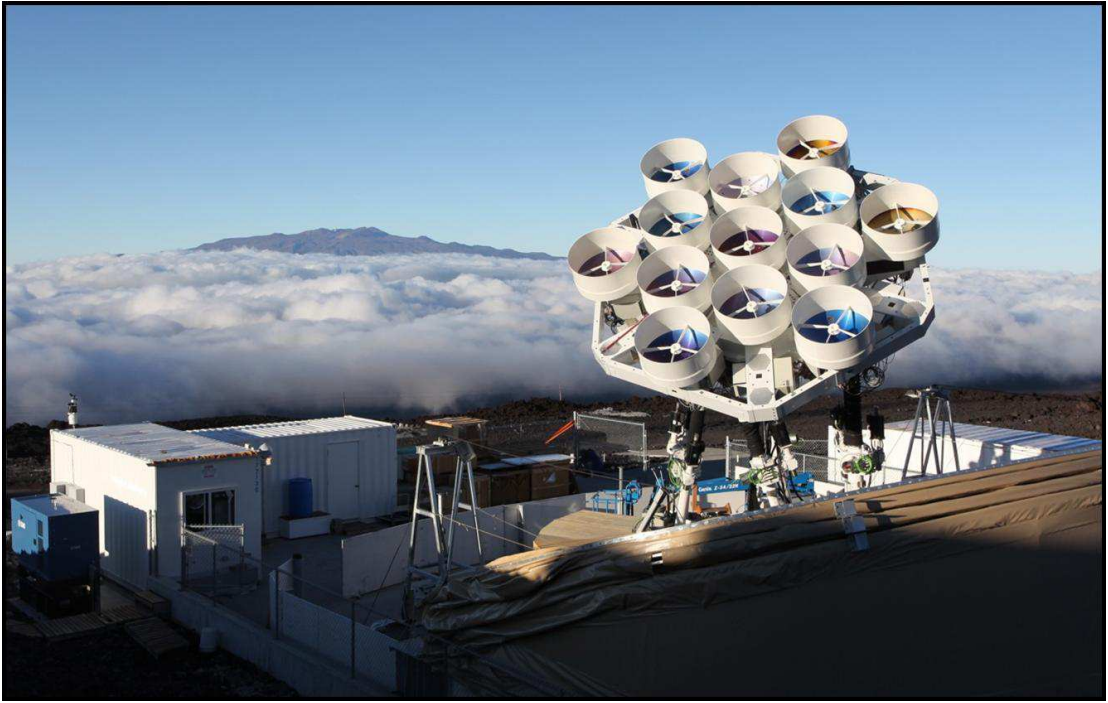}
\caption{\label{front}Front view of the AMiBA in the completed expansion phase 
with 13 1.2~m diameter Cassegrain antennas.
All antennas are co-mounted on a fully steerable six meter carbon-fiber
platform which is driven by a hexapod telescope.
The antenna shielding baffles are clearly visible.
Free receiver holes in the platform 
(covered with aluminum plates on the photo)
allow for different array configurations.}
\end{center}
\end{figure}

After an initial phase with seven 0.6~m diameter antennas \citep{koch06} in a compact configuration,
the AMiBA is currently operating with 13 1.2~m diameter Cassegrain antennas (Figure \ref{front}).
This new antenna 
and its capabilities
are described here. 
Section \ref{design} lists the antenna requirements. In Section \ref{mechanical}
the mechanical and optical designs are detailed out,
including simulation 
results of the structure and the antenna-feedhorn system. Section \ref{measurement}
is devoted to the antenna verification measurements. 
The factors composing the antenna efficiency are estimated in Section \ref{efficiency}.
Section \ref{novelty} discusses the improved design features of the 1.2~m antenna
and the upgraded array operation.
Our conclusion is given 
in Section \ref{conclusion}.

Previous AMiBA progress reports were given in \citet{ho04,raffin04,li06,raffin06}. 
A project overview is given in \citet{ho08}. More details about the correlator and 
receiver can be found in \citet{li10} and \citet{chen08}. The hexapod telescope 
mount is introduced in \citet{koch08}.  
Observing strategy, calibration scheme and data analysis with quality
checks are described in \citet{lin08,wu08,nishioka08}.
First AMiBA science results from the 7-element array are presented in 
\citet{huang10,liao10,koch08b,liu08,umetsu08,wu08}.
A possible science case utilizing the 1.2~m antenna array configuration is 
outlined in \citet{molnar10}.

\section{Antenna Requirements} \label{design}

Antenna size and interferometric baselines are constrained by the required window functions 
sampling the scales 
on the sky, which are relevant for our target science. 
After the AMiBA initial phase with seven 0.6~m diameter antennas in close-packed 
configuration, 13 1.2~m diameter antennas are now installed, covering a baseline range
from 1.4~m to about 6~m. This gives a synthesized beam resolution of about 2 arcmin
for a nominal central frequency of about 94~GHz
The collecting area of the 13-element array is increased by a factor of 7.4 upon 
the initial phase.

The antenna pointing accuracy and its mechanical stiffness requirements against wind force, 
self-gravity and thermal load are driven by its field of view (FoV), which is about $11^{\prime}$ 
at Full Width Half Maximum (FWHM) at 
94~GHz 
(Section \ref{simulation} 
and Section \ref{section_beam}).
We consider deformations and mechanical alignment errors resulting in less than a $1^{\prime}$ 
tilt ($\sim 10$\% of FWHM)
as acceptable. 
Minimizing deformations also ensures that the asymmetrical patterns on the antenna surface and the 
resulting antenna cross-polarization are kept at a low level. 
The manufacturing 
accuracy of the mirror surfaces is specified to be better than $50 \mu$m root-mean-square (rms) 
to ensure an antenna
surface efficiency of more than 95\% (Section \ref{cassegrain} and Section \ref{surface}).

Measuring the weak CMB fluctuations ($\sim 10 \mu$K) requires 
a low-noise antenna with low side lobe levels and very little scattering and cross talk.
High sidelobes are suppressed with a $-10.5$~dB edge taper in the feed's Gaussian illumination
pattern (Section \ref{simulation} and Section \ref{efficiency}). 
The antenna is shielded with a baffle to minimize the 
inter-antenna coupling and the ground pickup. Additional care is taken to send stray-light back
to the sky with  triangular roof-shaped quadripod legs (Section \ref{baffle_leg}).

Due to the harsh volcanic environment together with the large high-altitude temperature 
variations we decided to add a TiO$_2$ protection layer on both the primary and secondary mirrors
against abrasion. Furthermore,  
visual and infrared sun light is a potential hazard to burn our antennas. 
The 0.6~m diameter antennas were therefore additionally covered with a Gore-Tex layer on top
of the shielding baffle in order to absorb the damaging radiation.
At our observing microwave frequencies both the protection layer and the cover are designed to
minimize the absorption loss and maximize the reflectivity and the transmission, respectively 
(Section \ref{material}).
On the 1.2~m diameter antennas no Gore-Tex cover is needed
because most of the visible and infrared light is not reflected by the TiO$_2$ coated surfaces.

Finally, minimizing any torque and possible tilting of the antennas when mounted 
on the receivers is crucial
to reach a stable radio alignment. On top of that, in order to keep the total weight below the
acceptable limits of the hexapod telescope, a lightweight structure is 
needed together with the desirable radio 
frequency properties. Choosing carbon fiber reinforced plastic (CFRP) 
satisfies these conditions and keeps the weight of the 
antenna below 25kg.

\section{Mechanical and Optical Design}  \label{mechanical}

Section \ref{cassegrain} describes the basic Cassegrain antenna design.
The results from the antenna structure and 
antenna-feedhorn
simulations are described in Sections \ref{simulation_structure} and \ref{simulation}. 
The Sections \ref{baffle_leg} 
and \ref{feed_leg} 
discuss in more details the additional 
measures taken to 
control antenna cross-talk and to
reduce ground pick-up and stray-light.

\subsection{Cassegrain Antenna Geometry} \label{cassegrain}

\begin{figure}
\begin{center}
\includegraphics[scale=0.6]{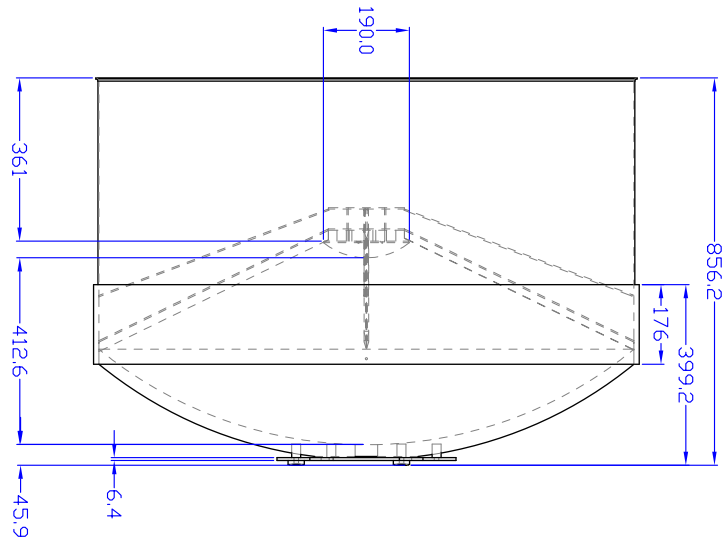}
\caption{\label{dish120_cut} Drawing of the assembled antenna including the main components:
primary and secondary mirror, structure and shield baffle, quadripod support legs and hexagonal
bottom support plate.
The units are in mm.
}
\end{center}
\end{figure}

\begin{figure}
\begin{center}
\includegraphics[scale=0.5]{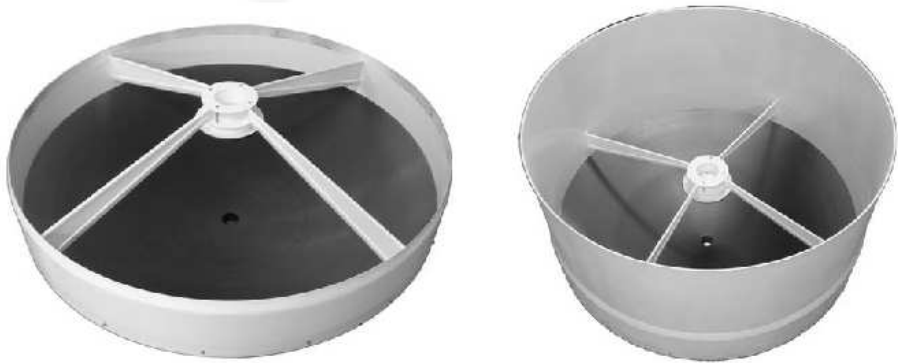}
\caption{\label{dish_picture} The AMiBA 1.2m Cassegrain antenna: on the left hand side the antenna with 
the structure baffle only, on the right the complete antenna with the shielding baffle.
The secondary mirror quadripod leg support is attached to the structure baffle.
}
\end{center}
\end{figure}

Based on requirements (Section \ref{design}), a shielded
f/0.35 Cassegrain antenna was chosen. 
Tables \ref{specs_primary} and \ref{specs_secondary} present the specifications for the 
primary paraboloid and the secondary hyperboloid mirrors. The Figures \ref{dish120_cut}
and \ref{dish_picture}  show a drawing and a picture of the 
assembled antenna, respectively.

The Cassegrain geometry  sets the feed phase center at the vertex 
of the primary. 
A parabolic illumination grading leads to  
about -20 dB for the first side lobe and a 11$^{\prime}$ FWHM for the 
primary beam at a wavelength of 
$\lambda \sim 3.2$ mm ($\sim$ 94~GHz).
Sub-reflector and feed positioning requirements are based on the Ruze formulas 
\citep{ruze66}.
 An axial and lateral secondary defocus of  0.1 $\lambda$ and 0.45 $\lambda$, respectively,  
keeps the gain loss at less than
 1 \%. Similarly, a feed horn positioning within 1 $\lambda$ gives a 99 \% gain. 
Random surface deviations from a primary paraboloid  
and a secondary hyperboloid  will remove power from the 
main beam and distribute 
it in a scattered beam. A surface error of less than 50 $\mu$m rms from the ideal geometry
ensures a 95\% gain.

\subsection{Structure Simulations}    \label{simulation_structure}

Typical load cases on the AMiBA site include wind, low temperatures
and gravity load.
Based on stiffness requirements and weight considerations, 
carbon fiber reinforced plastic (CFRP) was 
chosen for the dish, the feedlegs and the baffle. 
The primary and secondary mirrors are sandwich composites.
The baffle is made of 
two parts (Figure \ref{dish120_cut}):
a structural baffle supporting the feed-legs and a non-structural 
(shielding) baffle 
fixed to the latter to prevent cross-talk between antennas
and minimize ground pick-up.
Mass is an issue as the AMiBA platform is designed 
to accommodate up to 19 identical antennas, but the hexapod drive systems sets
an overall weight limit.
From an original 50 kg prototype antenna, the weight was reduced to about 
25 kg with the help of a Finite Element Analysis (FEA), 
while maintaining the overall structural behavior 
of the antenna.
The main goal of the structural analysis is to ensure that
the parabolic characteristics of the primary mirror are met for 
a minimized mass.

\begin{figure}
\begin{center}
\includegraphics[scale=0.5, angle=0]{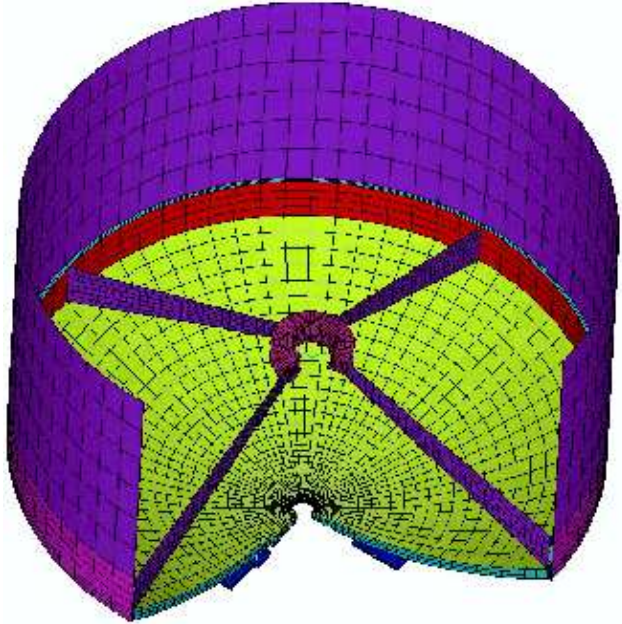}
\caption{\label{dish_grid}
Grid for the FEA antenna model.
15,400 nodes and 18,300 elements are used for the 
structure simulation.
The main components of the antenna are shown in different 
colors.
}
\end{center}
\end{figure}

\begin{figure}
\begin{center}
\includegraphics[scale=0.78, angle=-90]{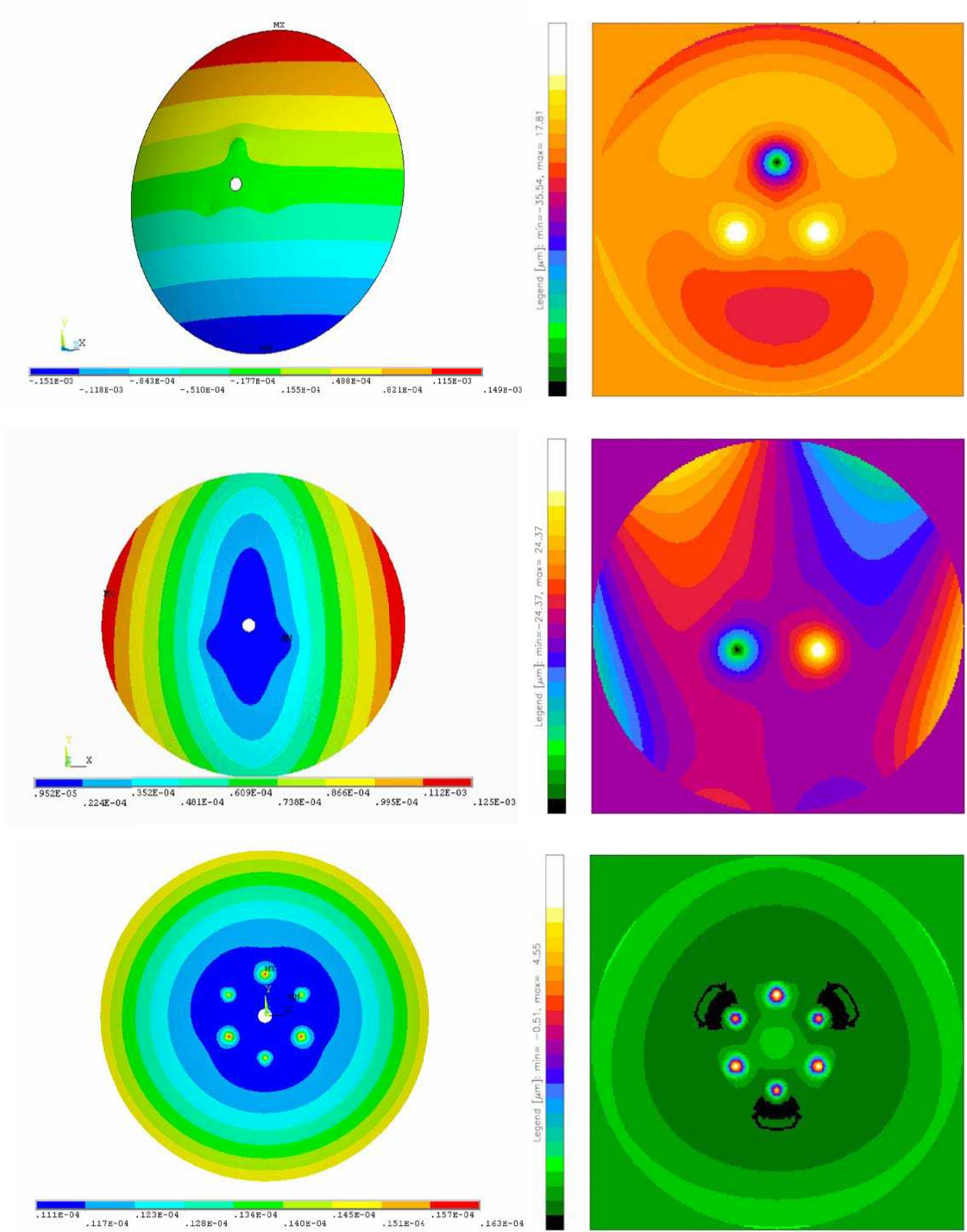}
\caption{\label{load_cases} \scriptsize
Selected load cases from structure simulations. The left panels display the 
deformation of the primary mirror. The units on the color bars are in $\mu$m. 
The corresponding residual maps from a best-fit paraboloid
to the deformed surfaces are shown in the right panels. 
The units are in $\mu$m.
The top panels illustrate the deformation due to gravity load along the $y$-axis
at an elevation of 30 deg. The maximum peak-to-peak (PTP) deformation is 0.3~mm (right panel), 
which leads to a 53 $\mu$m PTP and 4 $\mu$m rms deviation from a best-fit paraboloid (left panel).
A 10~m/s side wind load at $el=90$~deg is analyzed in the middle panels. 
The primary mirror maximum PTP normal deflection is 0.115~mm. Residual maximum PTP
and rms deviations are 49 $\mu$m and 5 $\mu$m, respectively. 
A uniform temperature increase of 20$^{\circ}$C is simulated in the bottom panels, 
leading to maximum normal deflections of 5 $\mu$m, and a residual maximum PTP
and rms deviation of 5 $\mu$m and 1 $\mu$m, respectively. 
}
\end{center}
\end{figure}

\begin{figure}
\begin{center}
\includegraphics[scale=0.5]{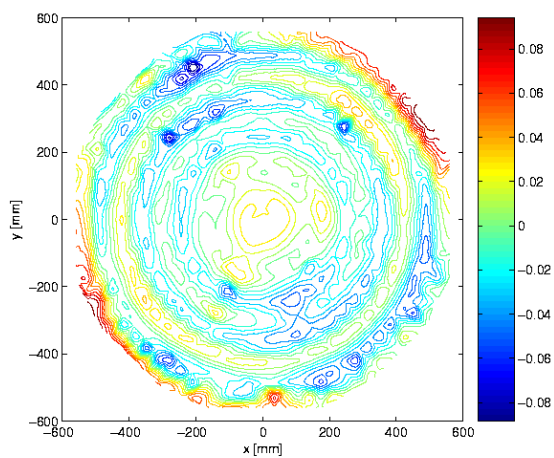}
\caption{\label{primary_contour} 
Contour plot of the fitting residuals from a a best-fit parabolic primary mirror. 
The units are in mm. The surface 
accuracy is about 28 $\mu$m rms, with maximum deviations of 96 $\mu$m and -98 $\mu$m, respectively, leading 
to a focal distance $F_p=413.66$~mm. The systematic ring-like pattern in the residuals reveal an imperfect
manufacturing process 
(likely resulting from an uneven support during the machining)
at a level which is irrelevant for our frequency range around 
94 GHz.
}
\end{center}
\end{figure}

\begin{figure}
\begin{center}
\includegraphics[scale=0.5]{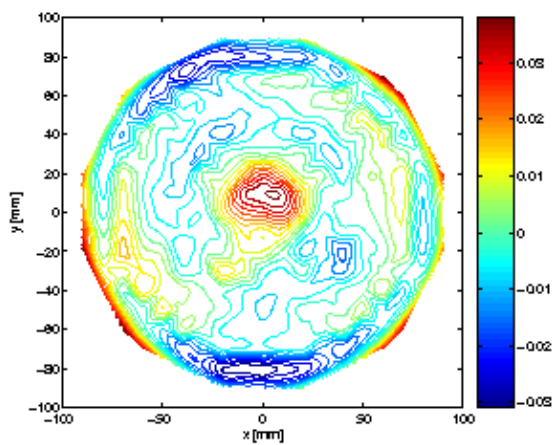}
\caption{\label{secondary_contour} Contour plot of the fitting residuals from a a best-fit hyperbolic 
secondary mirror. 
The units are in mm. The surface 
accuracy is about 13 $\mu$m rms, with maximum deviations of 42 $\mu$m and -34 $\mu$m, respectively, leading 
to a focal distance $F_p=413.59$~mm.}
\end{center}
\end{figure}

The simulation results presented in this section refer to the 
production antenna. Figure \ref{dish_grid} shows the grid of the FEA model.
The finite element model was analyzed with ANSYS version 11 with 
15,400 nodes and 18,300 elements. In particular, 
3D-Shell elements were used to model the CFRP primary mirror skins,
 the baffle and feed-legs, and the sub-reflector coupling.
3D-Solid elements were adopted to model the core of the primary mirror, 
the structural baffle and the hexagonal support plate sandwich composites.
Finally, 3D-Beam uniaxial elements served to model the sub-reflector and the inserts.
Boundary conditions apply on the hexagonal support plate and 
are not symmetrical with respect to the antenna geometry. 
Therefore, structural behavior varies with the antenna orientation in its aperture 
plane. For each individual load case (gravity, wind load and thermal load)
the deformation (Figure \ref{load_cases}, left panels) 
is calculated from a displacement vector which is 
the sum of the nodal components. The residual map from a best
fit paraboloid to the deformed primary mirror is shown in the 
right panels in Figure \ref{load_cases}.
Table \ref{summary_structure_simulation} summarizes the results.
For combined load cases the added surface deformations are within 
about 5~$\mu$m over the entire
elevation range. 
This only adds insignificantly to the manufacturing random errors 
($\epsilon_{\|} < 30 \mu$m, Section \ref{efficiency} and 
Figures \ref{primary_contour} and \ref{secondary_contour}) and, 
therefore, does not further reduce the efficiency.
Combined resulting tilts of the optical axis are within 1~arcmin, 
which is at the acceptable 10\% level of the FWHM ($\sim 11$ arcmin) around the 
operating frequency of
94~GHz.
This is comparable to the achievable mechanical alignment between
individual antennas on the platform which then keeps the loss in amplitude
for each antenna pair at the percent level.

\subsection{Feedhorn - Antenna Simulations}   \label{simulation}

\begin{figure}
\begin{center}
\includegraphics[scale=0.6]{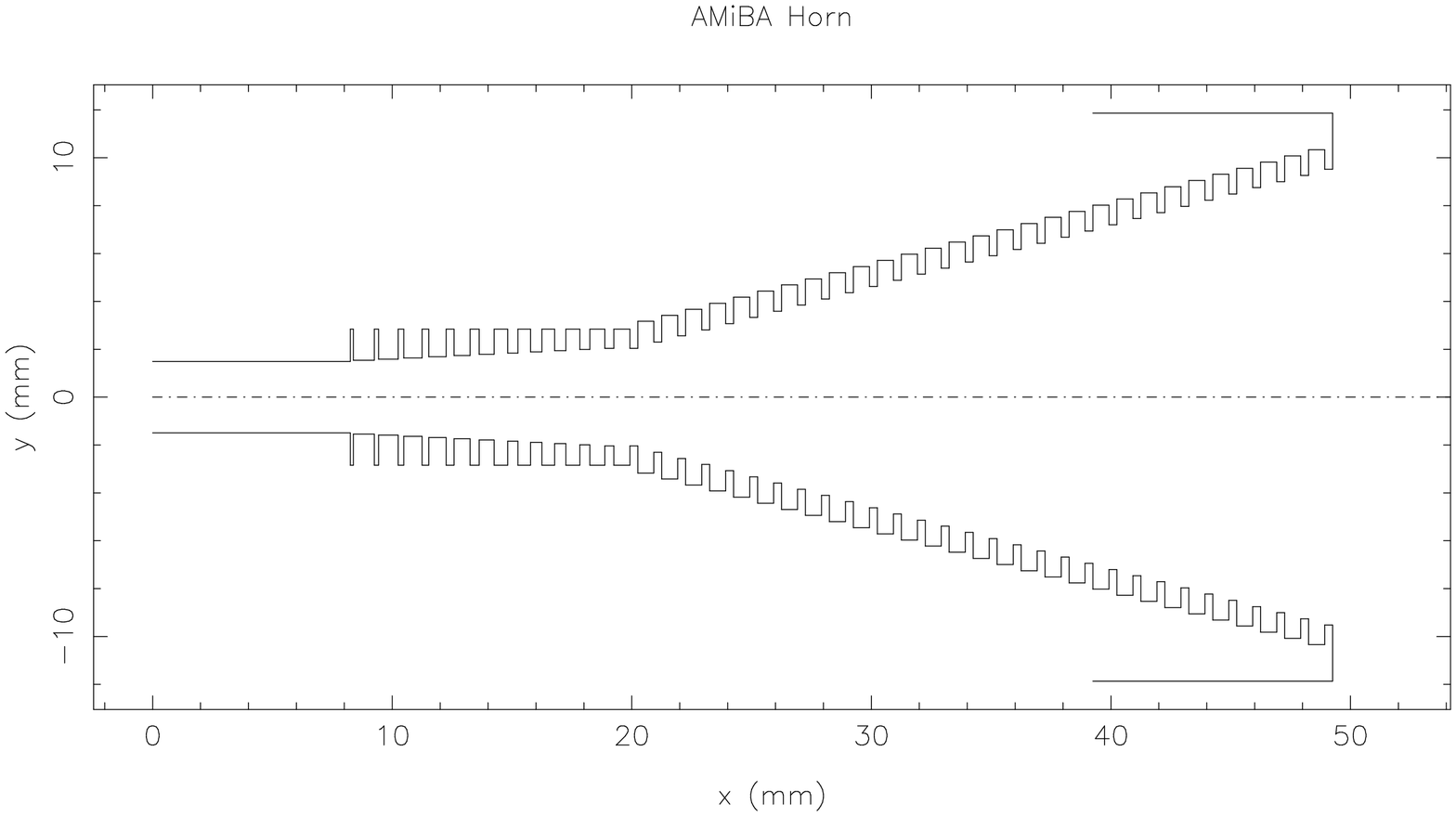}
\caption{\label{amiba_horn} Geometry of the AMiBA 85-105 GHz feedhorn.}
\end{center}
\end{figure}

\begin{figure}
\begin{center}
\includegraphics[scale=0.6]{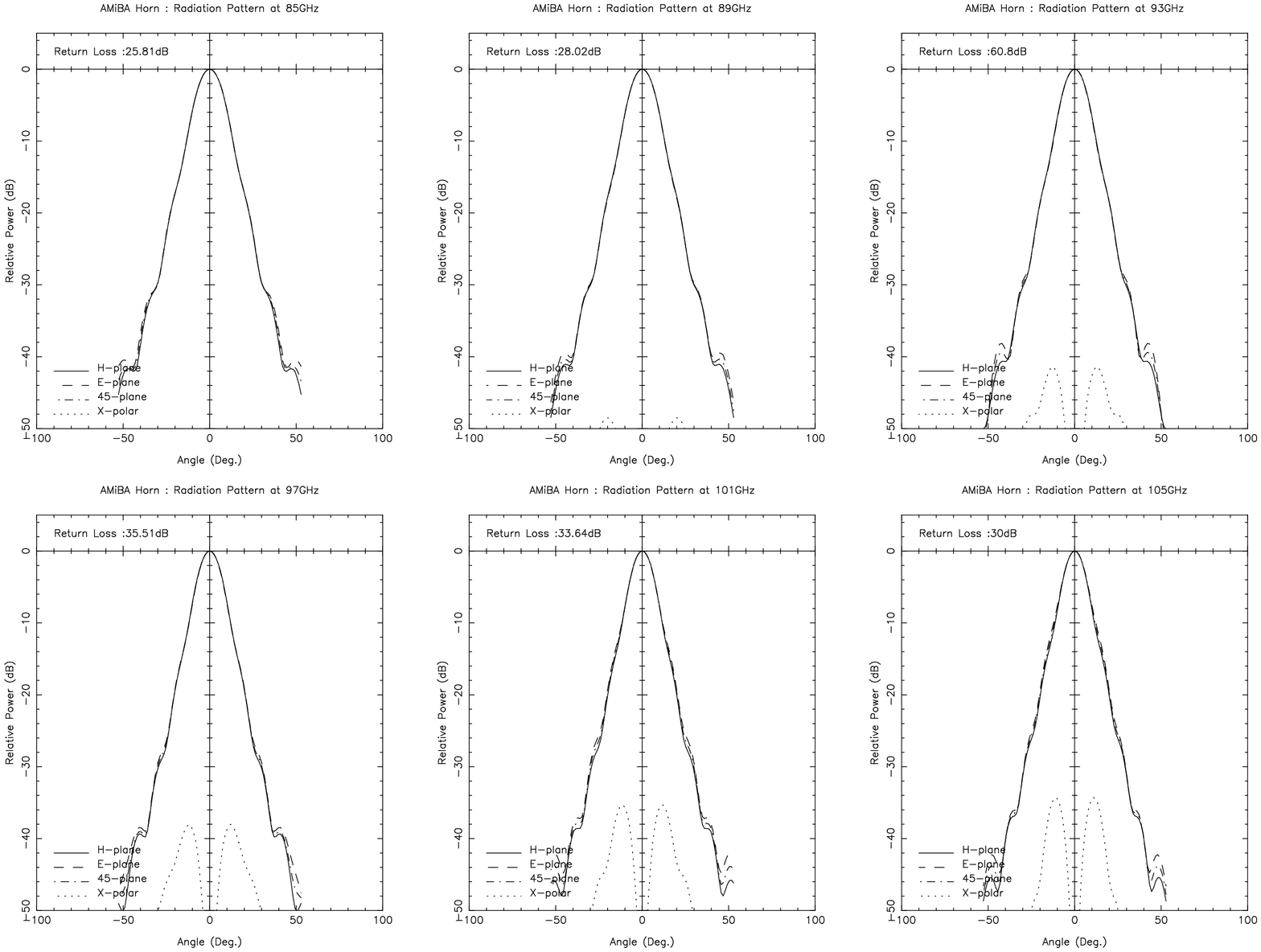}
\caption{\label{amiba_horn_pattern} Simulated radiation patterns of the AMiBA feedhorn 
over the 85-105 GHz frequency band.}
\end{center}
\end{figure}

We designed a corrugated feedhorn with a variable-depth mode converter to cover a
full 20~GHz band from 85 to 105 GHz
\citep{zhang93,granet05}. The center frequency is around 95~GHz.
The geometry of the horn is shown in Figure \ref{amiba_horn}. 
The feedhorn has a semi-flare illumination angle of 14$^\circ$
with a parabolic illumination grading with a -10.5 dB edge taper.
Free-space tapering for f/0.35 is about 3.6 dB, which leads to 
about -20 dB for the first side lobe and a 11$^{\prime}$ FWHM for the
primary beam at a wavelength of 
$\lambda=3.2$ mm.
The analysis of the horn is done using 
the well-proven mode-matching technique \citep{james81}. The resulting theoretical radiation pattern 
is shown in Figure \ref{amiba_horn_pattern}. An aperture flange with an external radius of $11.86$~mm
is used in the calculations. The phase-center of the horn is located 19~mm inside the horn, 
when measured from the aperture. The aperture of the horn is therefore located 19~mm above the 
vertex of the primary mirror.

The radiation pattern of the antenna has been simulated \citep{james2000}, using the following two assumptions: 
No feed-leg blockage and no baffle is taken into account. 
The radiation pattern of the horn, calculated using the mode-matching method, is then used to excite 
currents on the secondary mirror and the Physical Optics (PO) method is used to calculate the currents 
generated by the secondary onto the primary mirror. The radiation pattern of the antenna is then calculated 
using the combination of the contributions of the radiated power by the primary, 
secondary mirror and the feed \citep{james2000}.

\begin{figure}
\begin{center}
\includegraphics[scale=0.9]{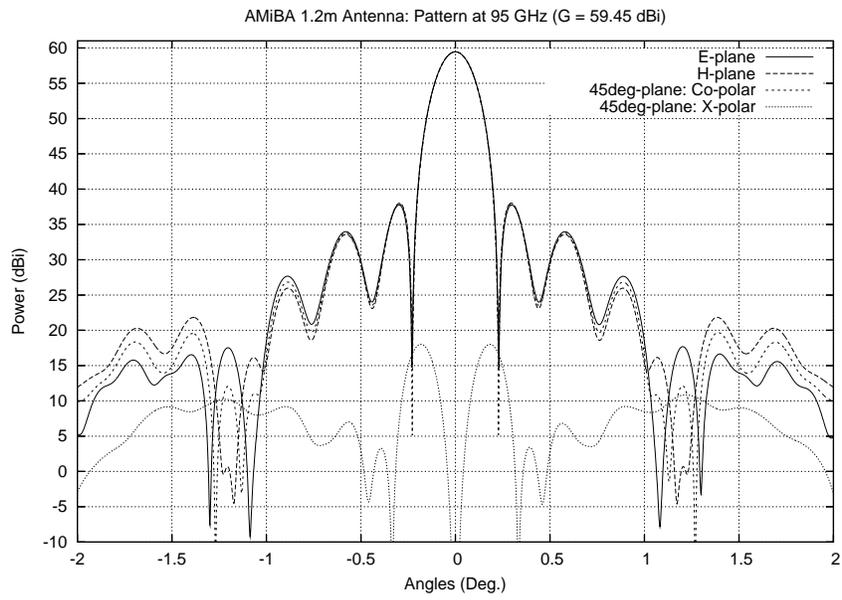}
\caption{\label{pattern_95GHz} Simulated radiation pattern at 95 GHz of the AMiBA 1.2m antenna.}
\end{center}
\end{figure}

The gain of the antenna is normalized by the input power of the horn, i.e., the return loss of the horn is 
included in the gain calculation. The efficiency calculation is based on a comparison between the 
gain of the antenna and the gain of an unblocked 1.2m-diameter with a constant amplitude and phase distribution.
The radiation pattern 
at $95$~GHz is shown in Figure \ref{pattern_95GHz} while the gain of the antenna over the $85-105$~GHz
band and the associated antenna efficiency are given in Table \ref{table_simulation}.

\subsection{Shielding Baffle}    \label{baffle_leg}

Having several antennas in a close-packed configuration on the 
platform can cause cross-talk problems which
might affect CMB measurements \citep{padin00}.
The total cross-talk signal $C_{tot}$ for a 
single antenna is expected to be proportional to the number of 
neighboring antennas. This leads to a maximum cross-talk
signal $C_{tot}\sim 6C$ for the central antenna in a hexagonal 
compact configuration (Figure \ref{front}), where $C$ is 
the coupling strength on the shortest baseline.
The cluster SZE signal is typically about 1~mK. The CMB temperature
fluctuations are $\sim 0.1$~mK 
on cluster scales and they decrease to  $\sim 10~\mu$K or less
on smaller scales.
In comparison, 
the maximum false signal due to cross-talk is $T_r\sqrt{pC}$, 
where $T_r$ is the receiver noise and $p$ is the correlation 
coefficient between the outgoing and receiver noises \citep{thompson82,padin00}.
Assuming $p=0.01$ and $T_r=65$~K for AMiBA \citep{chen08}, a maximum
wrong signal 
of $\sim 40\sqrt{C}$~K might be present at the central antenna. Adopting a 10\% 
tolerance of the weak CMB signal, 1~$\mu$K, $C$ should be reduced 
to about -127~dB.

Therefore, in order to assure a low inter-antenna cross-talk, it was 
decided to add a shielding baffle similar to the case of the 
Cosmic Background Imager (CBI).
Baffle height and baffle rim curvature are the important parameters
to determine.
Our design is closely following the antennas built for the CBI \citep{padin00}. 
Generally, the approach is
to eliminate sharp discontinuities which would diffract energy from 
the main lobe. Therefore, scattering from the baffle shield rim is 
minimized by rolling the rim with a radius $\sim 5\lambda\approx 15$~mm 
\citep{mather81}.

Finally, when using a shielding baffle, its effect on the 
forward gain of the antenna needs to be considered. As an approximation 
in the antenna near-field, the  Fresnel diffraction integral for a 
circular aperture is adopted to calculate the propagating electric 
field, $E(x,y,z)$, above the aperture plane. The aperture field 
distribution, $E(x,y,0)$, is a Gaussian with a -10.5 dB edge taper 
as a result of our feedhorn illumination. Evaluating $E$ in the center 
and at the radius of the baffle location as a function of distance 
from the aperture plane shows a close to constant ratio between them. 
This finding is consistent with the expected beam broadening, 
$w(z)=w_0\sqrt{1+(\frac{\lambda z}{\pi w_0^2})^2}$, along the optical 
axis, which is less than 0.1\% up to a height $z=1$~m for an initial
beam waist $w_0\approx 0.545$~m (for a Gaussian with a -10.5~dB edge
taper). Consequently, the power $P\sim E^2$ remains well confined 
with a plane integrated loss of less than 1\% (at a height of 1~m)
compared to a broadened beam without baffle.
The loss in forward gain does therefore not set any stringent constraint 
in this case due to the well collimated beam.
Eventually, it was decided to limit the baffle height to where 
the tangential plane at the edge of the secondary 
intercepts with the baffle. In the optical geometrical limit, 
rays scattered from the secondary are confined
to this plane. This results in a total baffle height of 
about $630$~mm above the aperture plane, which leaves the 
secondary and feedlegs about $360$~mm inside the baffle 
(Figure \ref{dish120_cut}).

\subsection{Secondary Mirror Feed Legs}  \label{feed_leg}

\begin{figure}
\begin{center}
\includegraphics{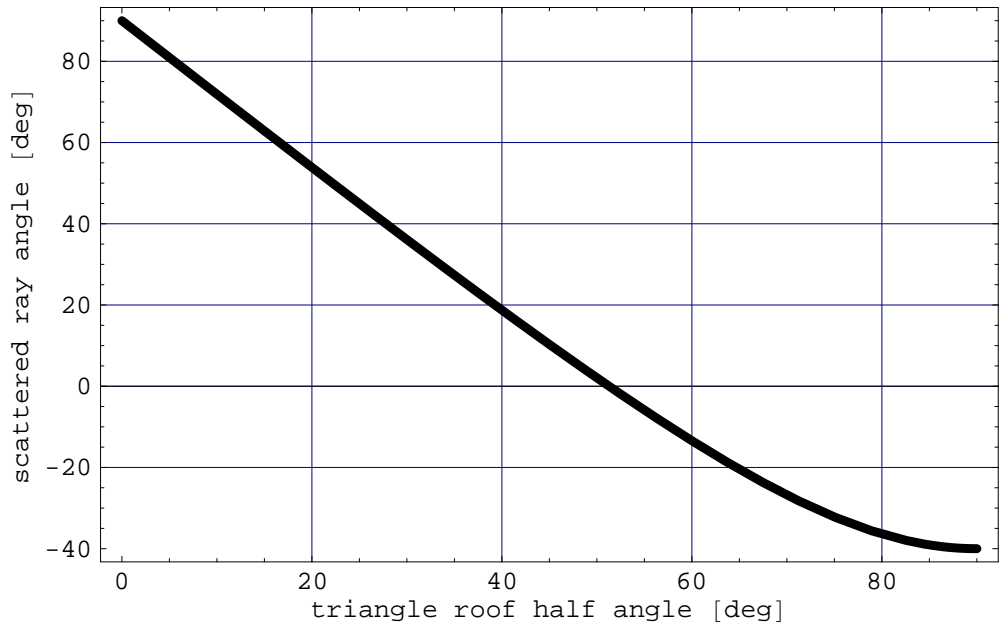}
\caption{\label{roofangle} Scattered ray angle ($\theta$) as a 
function of the triangular roof half angle ($\alpha$). The 
scattering angle is measured
with respect to the aperture plane. A positive angle means scattering
towards the sky, negative angle means scattering down towards the primary mirror. 
A triangular roof half angle of about $15^{\circ}$ was chosen.}
\end{center}
\end{figure}

Besides minimizing the antenna cross-talk, in order to further 
control the antenna sidelobe levels and scatter 
stray-light back towards the sky, 
the geometrical shape of the secondary mirror feed legs has been optimized.
Mainly two concerns are related to the feed leg shape: 
the antenna sidelobe level can increase due to diffraction 
from the legs and the system temperature can increase due 
to reflected ground pickup. 
\citet{satoh84,mo96} and \citet{la94} have extensively studied the issue of 
ground spillover pickup and feed leg 
shaping or baffling. One shaping technique is to attach a 
triangular roof on the lower side of the feed leg \citep{lamb98}. 
An optimized angle has been worked out for our Cassegrain 
geometry in order to keep scattered rays close to the pointing
axis towards the cold sky and away from the secondary and primary mirrors.  
Using geometrical optics, the scattered rays are found to be within a cone 
with opening angle $\theta$ \citep{cheng98}:
\begin{equation}
\theta= \pi/2-2\arcsin(\sin(\beta)\sin(\alpha)),
\end{equation}
where $\alpha$ is the half angle of the triangular roof 
and $\beta$ is the angle between the lower side of the 
feedleg and the optical axis of the antenna. For our design,
 $\beta = 65^{\circ}$, which is close to half of the 
primary mirror illumination angle because the feedlegs 
are attached close to the rim of the primary mirror (Figure \ref{dish120_cut}).
With this resulting scattering angle ($\theta$),  stray-light 
is terminated to the cold sky. The angle $\theta$ is also well 
matched with the baffle height in order to avoid multiple 
scatterings from the baffle, which might occur if $\theta$ is
too small and therefore intercepting with the upper parts of 
the shielding baffle. 
Since the height of the triangular roof increases with smaller $\alpha$, 
$\alpha \rightarrow 0^{\circ}$ - which would ideally send scattered rays
straight to the sky ($\theta\rightarrow 90^{\circ}$) - 
is practically not feasible. Therefore, as a compromise 
$\alpha\approx 15^{\circ}$ was chosen, which confines scattered rays
to within $\theta\approx 65^{\circ}$ (Figure \ref{roofangle})
and keeps the height of the equilateral roof shape within about 20~mm.
The 2-dimensional antenna beam pattern measurement in section \ref{section_beam}
possibly demonstrates an improvement due to this triangular roof shape.
Whereas for larger antennas an optimized leg design was able to reduce
the system temperature of up to 10~K, in the case of our small antennas
this might eventually be more a measure of precaution.

\section{Antenna Verification}     \label{measurement}

\subsection{Surface and Alignment Measurements}   \label{surface}

The mirror surfaces are measured at the Center for Measurement Standards (CMS), founded by the 
Industrial Technology 
Research Institute (ITRI) in Taichung, Taiwan, using a ZEISS PRISMO 10 measuring machine with a 
 4.4 $\mu m$ measurement accuracy.
The surfaces are checked from a sample of 
data points: 885 and 276 uniformly distributed points are measured across the primary and secondary
mirrors, respectively. This yields sets of $(x,y,z)$-coordinates for a two-dimensional fit for the 
primary paraboloid and the secondary hyperboloid, 
respectively, using the following formula:
\begin{eqnarray}
primary: &z&=\frac{x^2+y^2}{4F_p}+C_p\\
secondary:&z&=\sqrt{a^2+\frac{a^2}{b^2}(x^2+y^2)}+C_s,
\end{eqnarray}
where $F_p$ is the focal distance, and $C_p$ and $C_s$ are off-sets in the $z$-direction
due to the measurement set-up. 
$a$ and $b$ determine the curvature of the hyperboloid with a resulting
focal distance $F_p=2\sqrt{a^2+b^2}$ (Table \ref{specs_primary} and \ref{specs_secondary}).
The primary and secondary mirrors typically have surface rms errors of about
30 $\mu m$ and 15 $\mu m$, respectively, as illustrated for one antenna in the Figures 
\ref{primary_contour} and \ref{secondary_contour}.
The resulting focal length $F_p$ is found to be within $0.1$~mm of the specifications. 
This reduces the dish surface 
efficiency by less than 2\%.
After verification of the surface, the antenna is assembled and the secondary
mirror is mechanically shimmed to meet the alignment criteria. 
The alignment is typically between 
$50\,\mu m$ and 
$100\,\mu m$ in the $x$, $y$ and $z$ translation directions.
The aperture efficiency (Section \ref{efficiency}) is therefore reduced 
by only less than 1\%.

\subsection{Antenna Beam Pattern}  \label{section_beam}

For precision cosmology an accurately measured antenna beam pattern is essential.
The beam convolution effect reduces the information of the observed signal 
on small angular scales which is of particular importance for the CMB 
power in the multipole space. Furthermore, defects in the beam pattern, such as an  
increase in the side lobes or a circular asymmetry, can affect the 
sensitivity limits \citep{wu01}. Besides confirming the theoretical expectations,
the detailed measurement and characterization of the antenna beam pattern 
makes it possible to use the exact beam response in the later science data
analysis if this should be needed. 

\begin{figure}
\begin{center}
\includegraphics[scale=0.48]{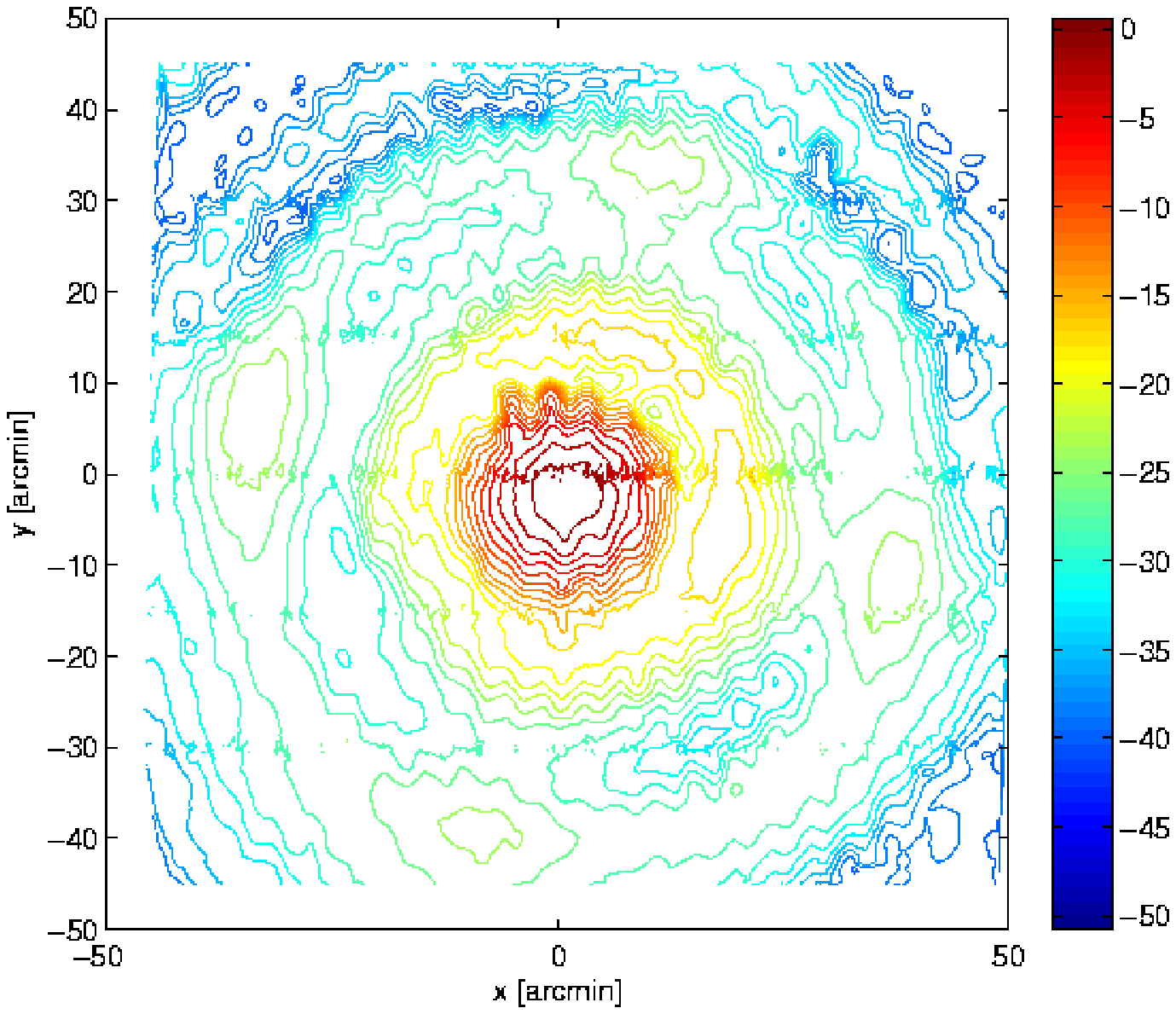}
\includegraphics[scale=0.54]{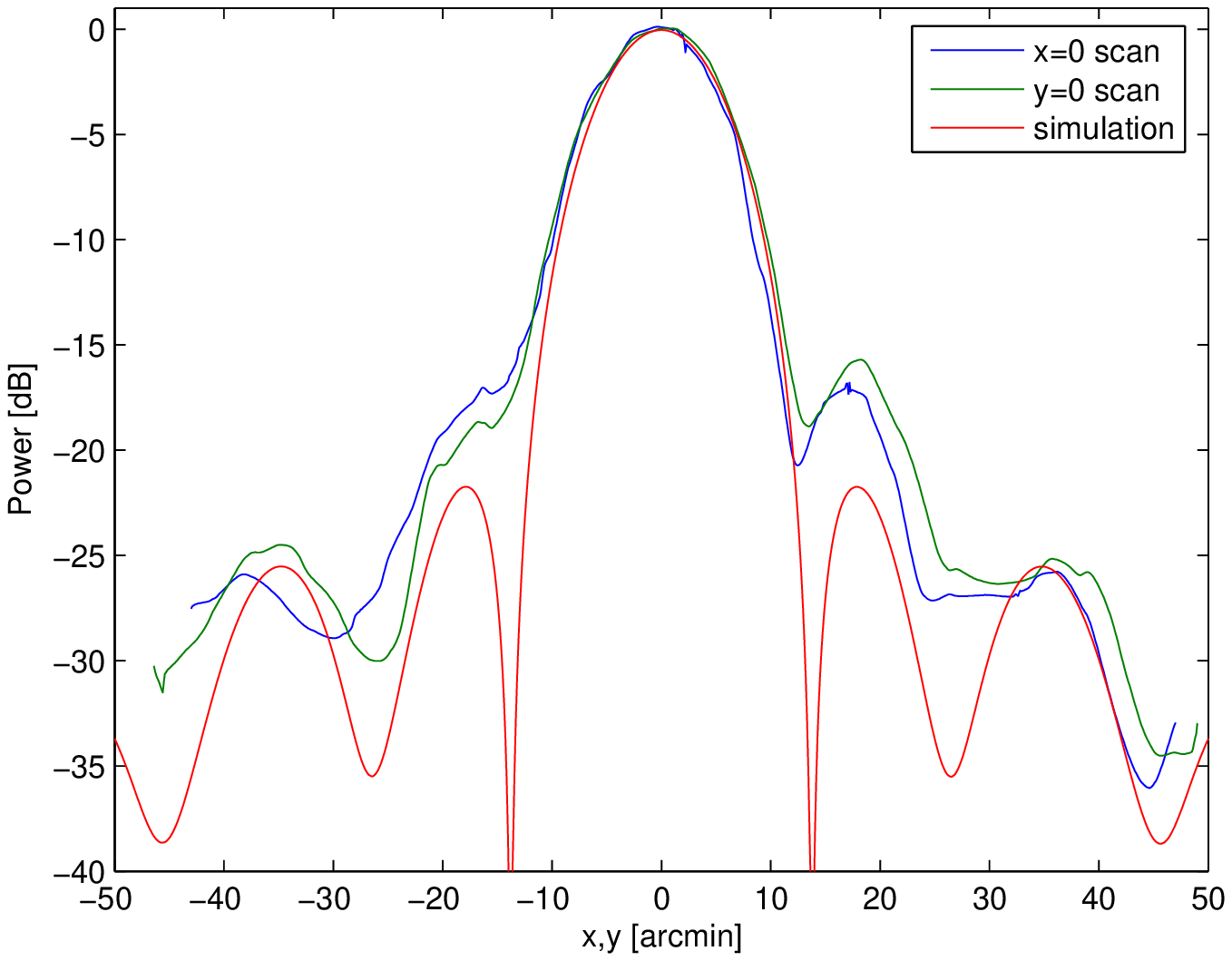}
\caption{\label{antenna_beam} 
Left Panel: Contour plot of the antenna beam pattern measured in the far. 
The units are in dB. 
The cross-structure around the secondary sidelobes is likely to result
from scattering of the secondary mirror support structure.
Right Panel: One-dimensional scans, extracted from the Left Panel, parallel to the $x$- and $y$-axis 
after location of the main beam maximum. Overlaid is the simulation result 
(section \ref{simulation}) for the E-plane at 95 GHz.}
\end{center}
\end{figure}

We measured the beam pattern with a computer-controlled equatorial mount, 
scanning a fixed 90 GHz thermally stabilized CW source 
at a distance of about $250-300$~m. This is marginally 
in the far field ($\sim D_p^2/\lambda \sim 480$~m).
An example of a measured beam pattern 
from a 2-dimensional scan
is shown in the Left Panel in Figure \ref{antenna_beam}. 
A cross-like feature likely resulting from the four legs of the secondary 
mirror support structure
becomes apparent, with an increase in power of $\sim 1$~dB at the location
of the second sidelobe. Comparing to the initial 0.6~m diameter antennas \citep{koch06} -
where a cross-like structure of a few dB was measured -  the feature has been reduced,
possibly with the help of the additional triangular roof at the secondary
support structure (section \ref{baffle_leg}).
The Right Panel in Figure \ref{antenna_beam} compares two orthogonal 
scans across the main beam center with 
the expected simulation result from section \ref{simulation} for the E-plane at 95 GHz.
Whereas the measured mainlobe nicely confirms the simulation result, the first sidelobe 
is about $2-4$~dB higher than expected.
The detailed cause of this is unclear. One might speculate that this is due 
to a weaker feed illumination edge taper than predicted and remaining stray light
from the secondary mirror support.
The location and the level of the second sidelobe agree again well with the simulation.
From the measurement, a FWHM of about 11 arcmin is found with
a first sidelobe peaking at about $-16$ to $-18$ dB.

The results are further analyzed
following \citet{wu01}, investigating azimuthally averaged beam profiles and $\omega_l$, 
the indices of asymmetry:
$\omega_l=(<|B_{lm}|^2>-<B_{lm}>^2)/<|B_{lm}|^2>$ . 
$ B_{lm}$ is the beam multipole expansion, $l$ being related to the corresponding angular scale. 
$<|B_{lm}|^2>$  and $<B_{lm}>^2$ are the mean of squares over $m$ and the square of the mean over $m$,
 respectively. 
The above defined numerator is thus the variance of $B_{lm}$ about its mean over $m$, 
a perfectly symmetric beam  
giving $\omega_l=0$. Calculating the indices of asymmetry  further requires the beam data to be pixelized. 
As a result, the index of asymmetry is always below 0.2 within the FWHM,
indicating a good symmetry of the antennas.

\subsection{Antenna Cross-Talk}  \label{cross_talk_measurement}

Low cross-talk signals were verified on the operating 13-element array on 
the AMiBA site. As a measurement set-up, one antenna served as an emitter
and was outfitted with a $\sim$ 10~dBm narrow-band polarized source
(Gunn oscillator with a bandwidth of $\sim$ 2-3~MHz at
the measured intermediate frequency (IF) of $\sim$ 5.56~GHz\footnote{
A measured IF frequency of  $\sim$ 5.56~GHz corresponds roughly 
to an original source radio frequency (RF) of $84+5.56=89.56$~GHz, with 
the receiver local oscillator frequency of 84~GHz. Since the source is not
in a thermally stable and cooled environment, some frequency drift ($\sim 1$~GHz)
in the RF is present. Such small variations are, however, irrelevant 
for the cross-talk measurments, and the source RF always remains in our 
observing band.
}
).
In different adjacent antennas (dual linear polarization receivers) 
the weak cross-talk signal 
was then detected with a spectrum analyzer connected to the output at 
the end of the first section in the IF chain.
Various amplifiers along the IF chain (from the receiving feedhorn
to the end of the first section) give rise to a gain of about 70~dB
\citep{chen08,li10}. Subtracting that and taking into account the source
power, the original cross-talk signal scattered into the feedhorn is 
derived.

\begin{figure}
\begin{center}
\includegraphics[scale=0.7]{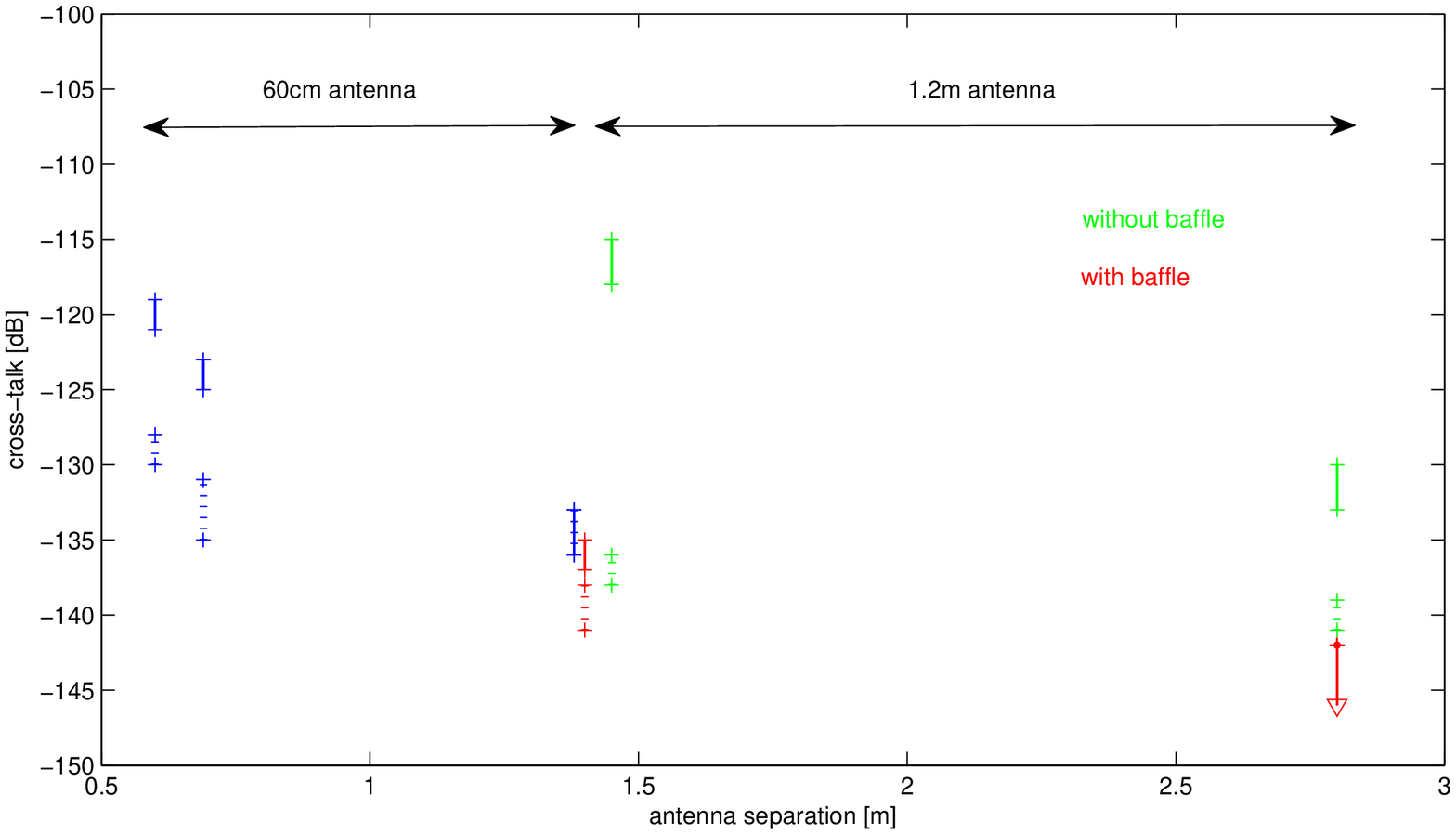}
\caption{\label{cross_talk_distance} Measured antenna cross-talk as a function of inter-antenna
distance. Results for the 1.2~m antennas are shown in red (with baffle) and in green 
(without baffle) for a 1.4~m and a 2.8~m baseline. 
For comparison also shown are the measured coupling strengths for the 60~cm antennas
for 0.6~m, 0.69~m and 1.38~m baselines (blue). 
(No removable baffles were constructed for those antennas.)
Solid and dotted lines are for maximally and 
minimally aligned polarizations, respectively. The down-arrow for the longest baseline
with shielding baffle indicates an upper limit.
Error bars include instrument noise and variations in the sky conditions.
For better display, measurements for the 1.4~m baseline are slightly shifted
horizontally.
}
\end{center}
\end{figure}

The coupling between two 1.2~m antennas was measured for a 1.4~m and a 
2.8~m baseline. In order to quantify the influence of the shielding 
baffle (Figure \ref{dish_picture}), the measurement was done with and 
without baffle. Figure \ref{cross_talk_distance} summarizes the results.
For comparison, the smaller 60~cm antennas were also checked on a few 
baselines. By varying the source orientation, cross-talk signals for  
maximally and minimally aligned polarizations were detected. For the 
1.2~m antennas without baffle, there is about a 20~dB difference on the 
shortest baseline depending on polarization. Generally, this difference 
is reduced for longer baselines, and it becomes indistinguishable on a
1.4~m baseline for the 60~cm antennas. When adding the shielding baffle, 
only a minor difference in polarization of $\sim$ 5~dB is found on the
shortest baseline for the 1.2~m antennas. The shielding baffle reduces
significantly the maximum cross-talk signal from about -115~dB to 
$\sim$ -135~dB or less on the 1.4~m baseline, and from $\sim$ -130~dB
to $\sim$ -145~dB or less on the 2.8~m baseline. As expected, for both 
the 60~cm and 1.2~m antennas, the cross-talk signal rapidly decreases
with distance. Thus, primarily the shortest baseline is of a concern.
Here, the shielding baffle successfully reduces the cross-talk below
the targeted  -127~dB level. For baselines beyond 2.8~m any remaining
coupling is beyond our detection limit of about -145~dB.

\subsection{Material Properties}  \label{material}

\begin{figure}
\begin{center}
\includegraphics[scale=0.7]{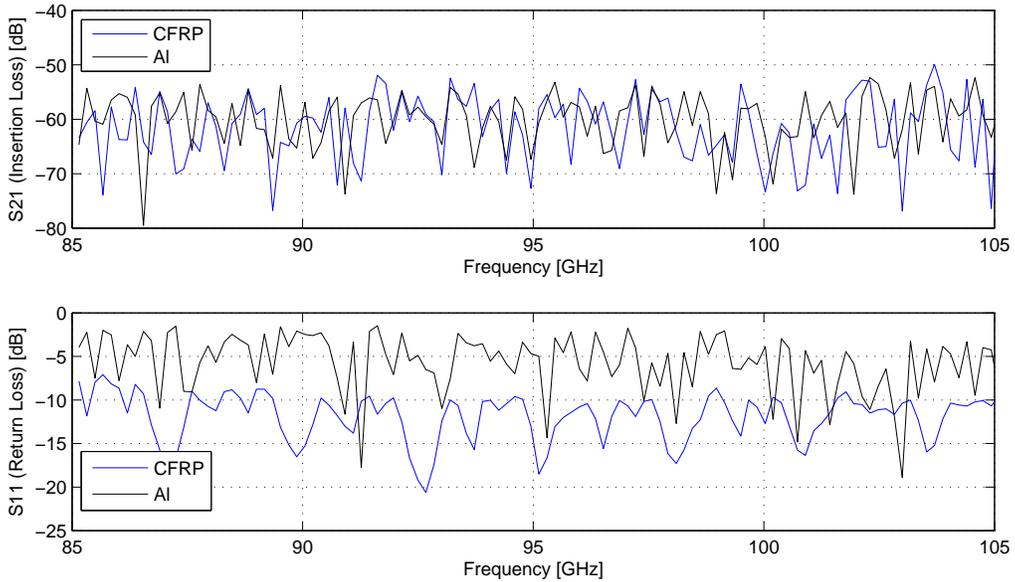}
\caption{\label{al_cfrp_s21}
VNA measurement: Upper Panel: Insertion loss (S21) for both CFRP and aluminum over the frequency range $85-105$~GHz. 
Lower Panel: Return loss (S11).
}
\end{center}
\end{figure}

Both the antenna structure and shielding baffle are made of carbon fiber reinforced plastic (CFRP).
The material has been measured with a vector network analyzer (VNA).
For a $0.72$~mm thick CFRP test sample an insertion (through put) loss (S21 parameter)
of about $-62$~dB or less is measured in the frequency range of $85-105$~GHz,
which is comparable to the $0.88$~mm aluminum
sample which shows an insertion loss around $-60$~dB (Figure \ref{al_cfrp_s21}, Upper Panel). 
The values are upper limits because they are at the noise level of the VNA.
The return loss (S11 parameter) averages around $-11$~dB in 
the observed
frequency range compared to about $-5$~dB for the aluminum sample (Figure \ref{al_cfrp_s21}, Lower Panel).
From this we conclude that very little radiation goes through the baffle and the deficiency in 
return loss (reflection) is absorbed in the material, which makes CFRP and ideal and lightweight
shielding material. 
Similar VNA tests have also shown that a possible Gore-Tex cover to protect 
the antenna additionally against visible and UV-light
has a transmission loss of less than $0.1$~dB.

The surface top layers of the primary and secondary mirrors are aluminum
coated in the vacuum. The reflectivity of aluminum is close to 100\% as 
soon as there are only a few layers of atoms. Further 
aiming at minimizing a possible emission from the underlying 
material, an aluminum layer attenuation to at least 1\% 
is targeted. A 5 times skin depth leads to a reduction in amplitude 
by a factor of 
$(1/e)^5=0.67$\%. This sets the coating layer
to $\delta=5\times\sqrt{2/(2\pi\nu\mu_0\sigma)}\approx$ 1.4 $\mu m$ 
at a frequency of 
$\nu=$ 94 GHz. 
$\sigma$ and 
$\mu_0$ are the  electrical conductivity for aluminum and the 
free space magnetic permeability, respectively.
 ($\sigma\approx 3.6\times 10^7 \Omega^{-1}$m$^{-1}$ and $\mu_0=1.2566\times 10^{-6}$Vs/Am.)
The typical aluminum layer - measured at various positions across the primary mirror - 
is about 2~$\mu$m\footnote{
For five specimen, four located at the outer radii and one in the center
of the primary mirror, the thickness of the Al$+$TiO$_2$ layer was measured with an
optical profilometer. From these test measurements, an average coating of $\sim 2\mu$m was derived, 
with slight variations probably resulting from different angles and distances to the 
sputtering source.
}
.

Additionally, a thin top layer of TiO$_2$ is applied for protection 
from oxidation, abrasion, peeling off and 
for thermal stability. 
Aiming for a thin layer of thickness $l$ ($l\ll \lambda/\sqrt{\epsilon}$) which is only very slightly
lossy ($\sigma/(\omega \epsilon)\le 0.01$), 
we set $l\sim 0.3\mu m$ ($0.001\times \lambda/\sqrt{\epsilon}$).
In here, we have adopted\footnote{
Thin films of deposited TiO$_2$ contain mainly two types of crystalline structures: 
Anatase and Rutile. The 
dielectric constant differs by about a factor of two ($86$~As/Vm and $173$~As/Vm) 
for parallel and perpendicular 
incoming waves (at $20^{\circ}$C and $\nu \sim$~1~MHz). No value for the conductivity $\sigma_{\rm TiO_2}$ 
at our operating frequency $\nu\sim$ 
94~GHz 
was found in the literature. Since adding O$_2$ increases
the conductivity  $\sigma_{\rm Ti}$, we use the available value, $\sigma_{\rm Ti}\sim 10^6$, 
for a conservative
estimation for $\sigma_{\rm TiO_2}$. A measured very small loss ($\ll 1\%$) for $\nu\sim 1-3$~GHz for a
$0.3\,\mu$m TiO$_2$ layer \citep{haney06} is supporting this estimate.
}
 $\epsilon\approx 80-170$ As/Vm and $\sigma\approx 10^6 \,\Omega^{-1}$m$^{-1}$ for the TiO$_2$ dielectric 
constant and electrical conductivity, respectively, at our operating wavelength 
$\lambda=3.2$ mm.
The criteria for a slightly lossy layer is easily met with $\sigma/(\omega \epsilon)\sim 10^{-8}$. 
This criteria also ensures that no microwave depolarization arises from the thin top layer 
\citep{chu76}.
The {\rm TiO}$_2$ vacuum sputtering is done immediately after the {\rm Al} sputtering. 
One sputtering run typically
gives a layer of about $0.15\mu$m.

We remark that this additional thin TiO$_2$ layer would even allow direct
sun observations. In tests where the antenna was directly pointed at the sun, 
a maximum temperature of 46$^{\circ}$C was measured after 20 minutes at the surface 
of the secondary mirror. The reflection of visible and infrared light is likely
to be significantly reduced due to multi-reflections in the TiO$_2$ layer.
Consequently, only a small portion of heat is reflected to the secondary 
mirror and further to the feedhorn. This small heat flux has also proven 
to be of no damage for the secondary mirror support structure made from CFRP.
For comparison, the initially used smaller 0.6~m diameter antennas
(without TiO$_2$ coating) showed a temperature increase to 70$^{\circ}$C
within only two minutes.

\section{Antenna Efficiency} \label{efficiency}

In order to derive an overall system efficiency for AMiBA \citep{lin08}, 
we analyze here the antenna aperture efficiency $\eta_a$. 
Generally, $\eta_a$ can be written as the product of a number of independent efficiency
components, e.g. \citet{kraus82}:
\begin{equation}
\eta_a=\eta_i \times \eta_b \times \eta_e \times \eta_s  \times \eta_f \times \eta_p \times \eta_c \times \eta_{misc} ,
\end{equation}
where:\\
$\eta_i$= illumination efficiency of the aperture by the feedhorn taper function\\
$\eta_b$= blocking efficiency due to the secondary mirror and its quadripod leg support\\
$\eta_e$= surface error efficiency due to small manufacturing random errors\\
$\eta_s$= spillover efficiency of the feed and the secondary mirror \\
$\eta_f$= focus error efficiency \\
$\eta_p$= cross-polarization efficiency of the feed-antenna combination \\
$\eta_c$= incoming radiation loss due to absorption from the Gore-Tex cover\\
$\eta_{misc}$= diffraction loss and reflector surface ohmic loss \\

Part of the following analysis is based on the guidelines in \citet{baars03}.
The illumination efficiency $\eta_i$ of the antenna is the ratio of the gain of the antenna to that of a 
uniformly illuminated aperture. An edge taper is applied to reduce the sidelobe levels. Our 1.2m 
antenna has a Gaussian illumination pattern with a $-10.5$~dB edge taper. 
For a Gaussian distribution, $F(r)=\exp(-\alpha(\frac{r}{r_0}))$, where $r_0$ is the aperture radius
and $\alpha=\frac{T_e}{20}\ln 10$ with $T_e$ the edge taper in dB, the illumination efficiency
$\eta_i=\frac{(\int F(r)\,dA)^2}{\int F^2(r)\,dA}$ becomes:
\begin{equation}
\eta_i=\frac{2(1-\exp(-\alpha))^2}{\alpha(1-\exp(-2\alpha))}.
\end{equation}
For our antenna we derive $\eta_i \approx 90$\%.

The loss of gain due to the blockage, $\eta_b$, is given by:
\begin{equation}
\eta_b=\left (1-\frac{A_b}{A_r}\right )^2, 
\end{equation} 
where $A_b$ and $A_r$ are the total blocked area and the aperture area, respectively. 
The double loss results from a decrease in the reflector area for the incoming plane wave front,
and a reduction of the incoming energy (spherical wave) to the focus. 
$A_b$ is composed by the central obstruction due to the sub-reflector ($A_{bc}$), the plane wave
shadow ($A_{bp}$) and the spherical wave shadow area. 
The spherical wave shadow (from the quadripod legs) is zero, since the legs are attached to the baffle
and not the primary reflector (Figure \ref{dish_picture}). 
Following \citet{baars03} for the case of a tapered illumination with a $-10.5$~dB edge taper, we have:
\begin{equation}
A_b=A_{bc}+A_{bp}=\pi R_s^2+\frac{4W_l*0.7}{3R_p^2}(R_p^3-R_s^3),
\end{equation}
where $R_p$, $R_s$ and $W_l$ are the radius of the primary and secondary mirror, and the width of the leg, 
respectively. 
The aperture surface $A_r$ is reduced by the apex hole. 
For our antenna geometry (Tables \ref{specs_primary} and \ref{specs_secondary}) 
we derive $\eta_b \approx 92.3$\%.

The surface error efficiency $\eta_e$ caused by small manufacturing random errors is calculated following
\citet{ruze66}:
\begin{equation}
\eta_e=\exp\left(-\left(\frac{4\pi\epsilon_{\|}}{\lambda}\right)^2\right),
\end{equation}
where $\lambda$ and $\epsilon_{\|}$ are the observing wavelength and the rms surface deviation parallel to the 
antenna axis, respectively. The manufacturing errors in $z$-direction with respect to the ideal geometrical
surfaces are illustrated in the Figures \ref{primary_contour} and \ref{secondary_contour}. They typically lead
to $\eta_e \approx 97.5$\%, ($\eta_e \approx 98.5$\% and $\eta_e > 99$\% for the primary and secondary 
mirrors, respectively), which  is significantly better than the commonly accepted error 
 $\epsilon_{\|}\approx \lambda/40$ which limits the gain loss to 10\%. 

The feed spillover efficiency $\eta_s$ measures the power radiated by the feed which is intercepted by 
the sub-reflector with the subtended angle $\Theta_s$ (Table \ref{specs_secondary}):
\begin{equation}
\eta_s=\frac{\int_0^{2\pi}\int_0^{\Theta_s}P_f(\theta,\phi)\sin\Theta\,d\Theta\,d\phi}
{\int_0^{2\pi}\int_0^{\pi}P_f(\theta,\phi)\sin\Theta\,d\Theta\,d\phi},
\end{equation}
where $P_f(\theta,\phi)$ is the power pattern of the feed.
From the secondary mirror illumination angle, $\Theta_s=28$ deg, and the feed illumination pattern, 
Figure \ref{amiba_horn_pattern} at $93$~GHz,
we derive $\eta_s\approx 0.78$.

The focus error efficiency $\eta_f$ is conveniently analyzed separately 
for secondary mirror and feed, and for axial
and lateral defocusing. Following \citet{baars03}, the gain loss for 
the sub-reflector axial defocusing $\delta_{z,s}$ 
can be expressed in units of wavelength $\lambda$ for different feed illumination patterns. 
For $\delta_{z,s}/\lambda<0.1$ (in our case $\delta_{z,s}/\lambda\approx 0.03$ for $\delta_{z,s}\sim0.1$~mm) 
the aperture
efficiency is better than $99$\% independent of the illumination pattern.
The axial defocus does not cause any change in the beam direction, and therefore there is no associated 
antenna pointing error.
The placement of the feed in the secondary focus is less critical \citep{butler03} 
with $\delta_{z,f}\sim \lambda$ still giving 
a $99$\% efficiency. We do not expect any efficiency loss here.
Similarly, the lateral defocus tolerances were analyzed in \citet{butler03}: a positioning error of less than 
$0.45 \lambda$ assures a loss less than $1$\%. 
The measured alignment (Section \ref{surface}) is well within this limit.
The resulting pointing error is 
less than $1^{\prime\prime}$. The sidelobe levels are unchanged with no Coma-lobes.

The instrumental polarization or polarization cross-talk is the ratio of the undesired orthogonal component
to the desired one. The effect can be minimized with precision parts: no asymmetries in mirrors and feedhorn, 
no squint feed and no antenna distortion.
Since the antenna reverses the polarization and since the primary and secondary surfaces have a parabolic 
and a hyperbolic curvature with a finite size, a parasitic cross-polarization component with the opposite 
polarization is introduced. The effect is curvature dependent and grows for a deeper dish.
We define the cross-polarization efficiency $\eta_p$ as:
\begin{equation}
\eta_p=\frac{\int_0^{2\pi}\int_0^{\pi}(P_C(\theta,\phi)+P_X(\theta,\phi))\sin\Theta\,d\Theta\,d\phi}
{\int_0^{2\pi}\int_0^{\pi}P_C(\theta,\phi)\sin\Theta\,d\Theta\,d\phi},
\end{equation}
where $P_C$ and $P_X$ are the co- and cross-polar patterns, calculated in Figure \ref{pattern_95GHz}.
Assuming a negligible contribution from asymmetries in the surfaces and the feed, 
the level of cross-polarization $P_X$
is more than $40$ dBi lower than the co-polarization within the main beam.
The resulting efficiency loss ($<0.01$\%) can therefore be neglected.

The finite sub-reflector size (based on a geometric optics design) 
produces diffraction causing further phase errors, amplitude taper losses
and cross-polarization changes. All these losses are summarized in a diffraction loss term. They have been 
analyzed numerically and the loss terms are given in a tabulated form in \citet{Lee79, milligan}. 
For our antenna parameters
the diffraction loss is about $0.22$~dB.
The efficiency term due to ohmic losses is usually negligible. Thus, $\eta_{misc}\approx 0.95$.

Finally, the absorption from a possible antenna Gore-Tex cover is about 0.1~dB 
(Section \ref{material}), and hence, the 
antenna cover efficiency $\eta_c\approx 0.98$.
All the efficiency components with the resulting aperture efficiency $\eta_a$ are summarized in 
Table \ref{efficiency_table}.

\section{Advancements and Comparison with Previous Array Operation}   \label{novelty}

\subsection{Antenna Design Novelty and Improvements}

From the initial 7-element array operation with 0.6~m antennas to the 
currently operating 13-element array with 1.2~m antennas several 
improvements were made. Table \ref{comparison_antenna} summarizes them
with their resulting benefits. Structure-wise, a major reduction in weight
per unit surface -- 10~kg for the 0.6~m antenna, $\le$25~kg for the 4 times
larger collecting area of the 1.2~m antenna -- without compromising on the 
antenna stiffness was achieved with detailed FEA modeling. This has been 
an important design requirement in order to keep the resulting torques on 
receiver and bracket within acceptable limits. In this way, a stable radio
alignment of $\sim 1^{\prime}$ between antennas is achievable. Maintaining
a mirror surface accuracy for the larger 1.2~m antennas to better than $\sim 30 \mu$m
leaves the possibility open to operate the array up to 300 or 400~GHz with 
possible future modifications. 

A significant improvement in the 1.2~m optical design was made possible
with a higher shielding baffle without reducing the forward gain of the
antenna. As a consequence, the inter-antenna coupling was reduced by more
than one order of magnitude as compared to the 0.6~m antennas to $\le -135$~dB.
Given sufficient integration time, this will allow for deep CMB observations
below the initial target level of 10~$\mu$K. Finally, a TiO$_2$ top layer is 
added on both primary and secondary mirrors as a protective measure. This 
makes the costly Gore-Tex cover redundant. At the same time, sun observations
for testing purposes are possible without damaging the mirrors. 

As a fundamental difference compared to earlier machined cast aluminum antennas
(CBI, \citet{padin00})
the AMiBA antennas are almost entirely made out of CFRP. The linear thermal
expansion coefficient $\alpha$  is about an order of magnitude smaller than
in the case of Al ($\alpha_{CFRP}\approx 1-2\times 10^{-6}$~K$^{-1}$,
$\alpha_{Al}\approx 23\times 10^{-6}$~K$^{-1}$). This reduces the thermal load
($\Delta T \approx 30$~K on the AMiBA site) and further ensures a very 
stable pointing of the antennas. The CFRP's superior tensile strength
($\sim 5000$~MPa compared to about 500~MPa for aluminum) combined with its
lower density ($\rho_{CFRP}\approx 1.2$~g/cm$^3$, $\rho_{Al}\approx 2.7$~g/cm$^3$)
make for a lightweight antenna ($\sim$ 25~kg). 
This is crucial in order to keep the overall
weight on the platform within the limits of the hexapod.
For comparison, an equally stiff antenna made out of aluminum is estimated 
to be at least 35~kg.

\subsection{Upgraded Scientific Capabilities}

A most immediate upgrade results from the significantly increased collecting 
area with the 1.2~m antennas.
The original seven 0.6~m antennas were deployed in a close-packed hexagonal pattern on 
the AMiBA platform in order to utilize the shortest spacings (0.6~m) of the interferometer.
This maximized the sensitivity for extended structures in clusters and 
primary CMB with a synthesized beam resolution of about 6 arcmin and a primary antenna
beam of about 22 arcmin FWHM. The point source sensitivity in 1 hour on-source integration
was about 63~mJy.
By replacing the 0.6~m antennas with  1.2~m diameter antennas, the collecting area 
and the speed in pointed observations are increased by a factor of 4 and 16, respectively.
The additional upgrade from 7 to 13 antennas leads to an overall speed-up 
factor of about 55. 
With the longest baselines a higher synthesized beam resolution of up to 2 arcmin
can be achieved. The resulting point source sensitivity is around 8~mJy/beam in 1 hour.
The primary antenna beam is reduced to about 11 arcmin FWHM.

Additionally, the larger number of antennas (78 baselines compared to 21 in the 
initial 7-element array) leads to a better $uv$-coverage with a higher dynamical 
range and, therefore, better imaging capabilities.
Whereas the initial array was only able to detect the cluster large-scale structures, 
the current 13-element upgraded array resolves weaker cluster substructures 
at the arcmin scale.
This will allow us to study more detailed cluster physics in combination with 
X-ray and optical data.
In continuous observation about 50 clusters can be detected per year.
In this way a substantial sample can be built up for statistical studies.
The CMB power spectrum can be probed up to a scale $l \sim 8000$, compared to the 
initial windows around $l \sim 1000, 2000$ and 2500.
The current science operations are focused on cluster observations.

\section{Summary and Conclusion}   \label{conclusion}

A 1.2~m $f/0.35$ Cassegrain antenna for a single platform close-packed 
interferometer for astronomical radio observations is presented. Due
to weight constraints, carbon fiber reinforced plastic (CFRP) is chosen 
as a lightweight material for the main antenna parts. With a detailed 
finite element analysis (FEA) it has been possible to keep the weight 
within 25~kg. The primary and secondary mirror sandwich composite structures
show excellent behavior under thermal, wind and gravity load, leading to FEA
predicted surface rms deformation errors of less than 10~$\mu$m and 
maximum tilts in the optical axis of about 1 arcmin. The primary 
paraboloid and secondary hyperboloid mirror manufactured surface rms 
errors are typically around 30~$\mu$m and 10~$\mu$m, respectively. The
mechanical alignment after shimming and the resulting focal length are
within $\sim 0.1$~mm of the specifications. The efficiency loss due to 
mechanical assembly and manufacturing is then within $\sim$ 1\%. For a
good reflectivity the mirrors are coated with a $\sim2\mu$m aluminum layer. 
Additionally, on top of that, a thin TiO$_2$ layer ($\sim 0.15\mu$m) 
protects the antenna from the harsh high altitude volcanic environment.

A corrugated feedhorn with a parabolic illumination grading with a 
$-10.5$~dB edge taper is used to achieve low sidelobe levels. The 
feedhorn antenna system is simulated and designed with the mode-matching
technique. The results are verified in a far-field beam pattern 
measurement. For the observing frequency around 
94~GHz, 
the first sidelobe
is around -20~dB with a main lobe FWHM of about 11 arcmin. With the goal
of sending more stray-light to the sky, legs with a triangular roof 
shape are added to the secondary mirror support structure. Despite this
attempt, a weak remaining feature in the beam map at the level of $\sim 1$~dB
around the secondary sidelobe is likely to be attributed to the 
secondary mirror support structure. 
Measuring the weak CMB signals ($\sim$ 10 $\mu$K) poses a challenge for 
a close-packed array due to inter-antenna coupling.
A CFRP shielding baffle is therefore added, 
which extends to a height of $\sim 360$~mm above the secondary mirror.
Insertion and return loss measurements show that CFRP is indeed an 
ideal lightweight shielding material. Without loss in the antenna forward
gain, the antenna cross-talk on the shortest separation of 1.4~m
is measured to be 
$\sim -135$~dB or less.

An overall antenna efficiency of about 60\% is estimated from a series 
of efficiency factors. The dominating loss results from the feed 
spill-over (efficiency $\approx 0.78$), followed by the illumination 
efficiency ($\sim 0.90$) and the secondary mirror blockage efficiency
($\sim 0.92$).

In summary, based on the calculated and measured properties, the presented
CFRP antenna is a lightweight, low side-lobe level and low noise antenna. 
Thus, it is appropriate for the targeted astronomical observations 
(Cosmic Microwave Background and galaxy cluster observations) in close-packed
antenna array configurations. Currently, a 13-element compact array is 
used in daily routine observations.

\acknowledgments

Capital and operational funding for AMiBA came from
the Ministry of Education and the National 
Science Council as 
part of the 
Cosmology and Particle Astrophysics
(CosPA) initiative. Matching operational funding also came 
in the form of an Academia Sinica Key Project. 
PMK acknowledges Shiang-Yu Wang for advising and consulting
on mirror coatings.

\clearpage

\begin{table} 
\begin{center}
\begin{tabular}{|l|l|}\hline

 Equation of Primary Mirror &  $z=(x^2+y^2)/(4F_p)$\\ \hline 
 Equation of Primary Illumination angle & $ \cot(\Theta_p/4)=4F_p/D_p$ \\ \hline 
 Primary Illumination Angle & $\Theta_p=142.151^{\circ}$ \\ \hline
 Diameter of Primary Mirror & $D_p=1182.000$~mm  \\ \hline
 Primary Focal Length & $F_p=413.700$~mm \\ \hline
 Depth of Primary Mirror & $H=211.071$~mm \\ \hline
 Primary f/ ratio  & $f/=F_p/D_p=0.35$ \\ \hline
 Apex hole diameter & $W=50$~mm  \\ \hline

\end{tabular}
\end{center}
\caption{\label{specs_primary}Primary mirror specifications.}
\end{table}

\begin{table} 
\begin{center}
\begin{tabular}{|l|l|}\hline

 Equation of Secondary Mirror &  $z=\sqrt{a^2+(a^2/b^2)(x^2+y^2)}$\\ \hline 
                              & $a=L(M-1)/2$ \\
			      & $b^2=L^2M$ \\
			      & $ c^2=a^2+b^2$ \\
			      & $L=2c/(M+1)=c-a$ \\
			      & $M=(c+a)/(c-a)$ \\ \hline
 Final Focal Position & at Vertex of Primary \\ \hline
 Distance between Prime and Secondary Foci  & $2c=F_p=413.700$~mm \\ \hline
 Final effective f/ ratio  &   $f/=2.0361$  \\ \hline
 Final Illumination angle & $\theta_s=28.000^{\circ}$ \\ \hline
 Magnification Factor & $M=5.81739$  \\ \hline
 Secondary Vertex to Prime Focus &  $L=60.683$~mm \\ \hline
 Equation Parameters & $a=146.167$~mm \\
                     &  $b=146.363$~mm \\ \hline
 Secondary Mirror Diameter & $D_s=2F_p/(\cot(\Theta_s/2)+\cot(\Theta_p/2))$ \\ 
                           & $D_s=190.048$~mm	\\   \hline  
 
\end{tabular}
\end{center}
\caption{\label{specs_secondary}Secondary mirror specifications.}
\end{table}

\begin{table} 
\begin{center}
\begin{tabular}{|c|c||c|c|c|c|c|}\hline

 load & el    & rms       & tilt     & $\Delta_x $ & $\Delta_y$  &  $\Delta_z$  \\ 
      & (deg) & ($\mu$m) & (arcsec) & ($\mu$m)   & ($\mu$m)   &  ($\mu$m)   \\ \hline\hline

gravity & 90  &  1       & 0        & 0           &  0          &     -9         \\
        & 60  &  3       & 40       & -59          &  0          &     -7         \\
        & 30  &  5       & 70       & -102          &  0          &     -4         \\ \hline
wind    & 90  &  5       & 63       & -93          &  0          &     0         \\
        & 30  &  2       & -27       & 39          &  0          &     -2         \\ \hline
thermal, 20$^{\circ}$C uniform & - & 1 &  0   & 0 &          0          & 11             \\
thermal, $\Delta \theta_x=2^{\circ}$C & - & $<$1 &  $<$1   & 0 &          0         & 0     \\
thermal, $\Delta \theta_y=2^{\circ}$C & - & $<$1 &  $<$1   & 0 &          0         & 0    \\
thermal, $\Delta \theta_z=2^{\circ}$C & - & $<$1 &  0   & 0 &          0         & 0    \\ \hline

\end{tabular}
\end{center}
\caption{\label{summary_structure_simulation}Summary of deformation from various load cases.
{\it el} refers to the antenna elevation position where $el=90$ deg means pointing at zenith. 
The wind load is for a 10~m/s side wind. The thermal load is independent of elevation. $\Delta \theta_i$
denotes a total temperature gradient in direction $i$, linearly distributed over the entire length.
The rms values refer to the residual maps after a best fit paraboloid is subtracted from the deformed
primary mirror surface.
Tilt values are different for $x$ and $y$-direction due to the non-symmetrical hexagonal mounting plate
of the antenna and, therefore, depend on the orientation of the antenna. 
Only the larger of the two cases is listed here.
$\Delta_i$ describes the shift in the primary apex in direction $i$. The antenna focal axis is in 
$z$-direction.
For combined load cases the added surface 
deformations are within about 5~$\mu$m rss (root of sum squares) over the entire
elevation range. Combined resulting tilts are within 1~arcmin.}
\end{table} 

\begin{table} 
\begin{center}
\begin{tabular}{|c||c|c|c|c|c|}\hline
Frequency [GHz]& 85   & 90   &  95  &  100  &  105   \\ \hline \hline
 Gain [dBi]     &  58.55 &  58.98 &  59.45 &  59.70 &  60.13 \\ \hline 
 Efficiency [\%]& 64.6  & 63.6 &  63.6  &  60.8 &  60.9  \\ \hline
\end{tabular}
\end{center}
\caption{\label{table_simulation}Antenna gain and efficiency over the $85-105$~GHz band.}
\end{table}

\begin{table} 
\begin{center}
\begin{tabular}{|c|c|c|c|c|c|c|c||c|}\hline

 $\eta_i $ & $\eta_b $ & $\eta_e $ & $\eta_s$  & $\eta_f$  & $\eta_p$  & $\eta_c$ & $\eta_{misc}$  & $\eta_a $ \\ \hline \hline
 0.90 & 0.923  &  $0.975^{\ast}$   & $<0.78$   &  $0.99^{\ast}$   & $\approx 1 $   & 0.98  & $<0.95$  &   $<0.58$ \\ \hline 

\end{tabular}
\end{center}
\caption{\label{efficiency_table}The individual estimated efficiency components and the resulting antenna aperture 
efficiency $\eta_a $. $\eta_s < 0.78$ is an upper limit because the feed power pattern is limited to the range 
$\pm\pi/2$. $\eta_{misc}<0.95$ is a conservative limit accounting for possible miscellaneous components. 
Values with an asterisk ($\ast$) are derived from measurements, others are calculated using a model.}
\end{table}

\begin{table}  \scriptsize
\begin{center} 
\begin{tabular}{|c|| c| c| c|}\hline
  & 1.2~m dish   & 0.6~m dish   &  benefit / comment  \\ \hline \hline
            &                 &                    &   \\ 
antenna             &                 &                    &   \\ \hline \hline
primary beam FWHM &   11 arcmin      & 22 arcmin      & -- \\ \hline
weight  &  $\le 25$~kg     & $\sim 10$~kg &  reduced mass per unit surface  \\ \hline
surface accuracy &       $\sim 30 \mu$m  & $\sim 30 \mu$m   & observations possible up to 300 - 400~GHz \\ \hline
surface treatment & TiO$_2$ coating  & none (Gore-Tex cover)  & better protection, sun observation \\ \hline
cross-talk   &  $\le -135$~dB         &  $\le -120$~dB         & improved sensitivity, reduced ground pick-up \\ \hline
secondary feed leg shape &  triangular roof  &  flat         & second sidelobe reduced by a few dB \\ \hline \hline
            &                 &                    &   \\ 
array             &                 &                    &  
                      $13\times 1.2$~m versus $7\times 0.6$~m, compact hexagonal  \\ \hline \hline
synthesized beam FWHM   &  2 arcmin    & 6 arcmin            &   natural weighting \\ \hline
array sensitivity     &    8 mJy/beam/hour            & 63 mJy/beam/hour  &   point source sensitivity\\ \hline
number of baselines  &   78      & 21   &   -  \\ \hline
individual baselines &   1.4; 2.42; 2.8; 3.7; 4.2; 4.84        &  0.6; 1.04; 1.2     &  -      \\ \hline
detection rate       &    $\sim$50 clusters per year        &  6 (total)    &   pointed observations    \\ \hline
\end{tabular}
\end{center}
\caption{\label{comparison_antenna}Improved 1.2~m antenna design, as compared to the initial
0.6~m antenna, and the resulting array characteristics. With the initial $7\times 0.6$~m array
 a total of 6 clusters was observed. }
\end{table}

\end{document}